\documentclass[11pt]{article}
\usepackage{epsfig}
\input epsf
\catcode`\@=11
\def\marginnote#1{}

\newcount\hour
\newcount\minute
\newtoks\amorpm
\hour=\time\divide\hour by60
\minute=\time{\multiply\hour by60 \global\advance\minute by-
\hour}
\edef\standardtime{{\ifnum\hour<12 \global\amorpm={am}%
    \else\global\amorpm={pm}\advance\hour by-12 \fi
    \ifnum\hour=0 \hour=12 \fi
    \number\hour:\ifnum\minute<100\fi\number\minute\the\amorpm}}
\edef\militarytime{\number\hour:\ifnum\minute<100\fi\number\minute}
\def\draftlabel#1{{\@bsphack\if@filesw {\let\thepage\relax
  \xdef\@gtempa{\write\@auxout{\string
    \newlabel{#1}{{\@currentlabel}{\thepage}}}}}\@gtempa
    \if@nobreak \ifvmode\nobreak\fi\fi\fi\@esphack}
     \gdef\@eqnlabel{#1}}
\def\@eqnlabel{}
\def\@vacuum{}
\def\draftmarginnote#1{\marginpar{\raggedright\scriptsize\tt#1}}
\def\draft{\oddsidemargin -.5truein
        \def\@oddfoot{\sl preliminary draft \hfil
        \rm\thepage\hfil\sl\today\quad\militarytime}
        \let\@evenfoot\@oddfoot \overfullrule 3pt
        \let\label=\draftlabel
        \let\marginnote=\draftmarginnote

\def\@eqnnum{(\theequation)\rlap{\kern\marginparsep\tt\@eqnlabel}%
\global\let\@eqnlabel\@vacuum}  }
\def\preprint{\twocolumn\sloppy\flushbottom\parindent 1em
        \leftmargini 2em\leftmarginv .5em\leftmarginvi .5em
        \oddsidemargin -.5in    \evensidemargin -.5in
        \columnsep 15mm \footheight 0pt
        \textwidth 250mmin      \topmargin  -.4in
        \headheight 12pt \topskip .4in
        \textheight 175mm
        \footskip 0pt

\def\@oddhead{\thepage\hfil\addtocounter{page}{1}\thepage}
        \let\@evenhead\@oddhead \def\@oddfoot{} \def\@evenfoot{}
}
\def\titlepage{\@restonecolfalse\if@twocolumn\@restonecoltrue\o
necolumn
     \else \newpage \fi \thispagestyle{empty}\c@page\z@
        \def\thefootnote{\fnsymbol{footnote}} }
\def\endtitlepage{\if@restonecol\twocolumn \else  \fi
        \def\thefootnote{\arabic{footnote}}
        \setcounter{footnote}{0}}  
\catcode`@=12
\relax
\newcommand{\newc}{\newcommand}
\newcommand\eg{{\it {e.g.}}}
\newcommand\ie{{\it {i.e.}}}

\newcommand\lsim{\mathrel{\rlap{\lower4pt\hbox{\hskip1pt$\sim$}}
    \raise1pt\hbox{$<$}}}
\newcommand\gsim{\mathrel{\rlap{\lower4pt\hbox{\hskip1pt$\sim$}}
    \raise1pt\hbox{$>$}}}

\newc{\gfermi}{G_F}

\newc{\sthw}{\sin\theta_W}              \newc{\cthw}{\cos\theta_W}
\newc{\bino}{\widetilde B}              \newc{\wino}{\widetilde W_3}
\newc{\higgsinob}{{\widetilde H}^0_b}   \newc{\higgsinot}{{\widetilde H}^0_t}

\newc{\abund}{\Omega h^2}
\newc{\abundchi}{\Omega_\chi h^2}
\newc{\rhocrit}{\rho_{crit}}
\newc{\rhochi}{\rho_{\chi}}

\newc{\mwimp}{m_{\chi}}     \newc{\rhowimp}{\rho_{\chi}}

\newc{\mplanck}{M_{\rm P}}              \newc{\mgut}{M_{\rm GUT}}
\newc{\mz}{m_{Z}}                       \newc{\mw}{m_{W}}

\newc{\ra}{\rightarrow}
\newc{\beq}{\begin{equation}}
\newc{\eeq}{\end{equation}}
\newc{\bea}{\begin{eqnarray}}
\newc{\eea}{\end{eqnarray}}

\newc{\vearth}{v_{\otimes}}
\newc{\vsun}{v_{\odot}}
\newc{\vsube}{v_e}	\newc{\vzero}{v_0}	\newc{\vrms}{v_{\rm rms}}

\newc{\sigmazero}{\sigma_0}
\newc{\sigmaa}{\sigma_A} 	\newc{\sigmaalim}{\sigma_A^{\mathrm{lim}}}
\newc{\sigmaap}{\sigma_A^p} 	\newc{\sigmaan}{\sigma_A^n}

\newc{\sigmap}{\sigma_p}	\newc{\sigman}{\sigma_n}
\newc{\sigmapa}{\sigma_p^A} 	\newc{\sigmana}{\sigma_n^A}
\newc{\sigmapap}{\sigma_p^{(A,p)}} 	\newc{\sigmanan}{\sigma_n^{(A,n)}}

\newc{\sigmaplima}{\sigma_p^{\mathrm{lim(A)}}}	
\newc{\sigmanlima}{\sigma_n^{\mathrm{lim(A)}}}

\newc{\sigmaplimai}{\sigma_p^{\mathrm{lim({A_i})}}}
\newc{\sigmanlimai}{\sigma_n^{\mathrm{lim({A_i})}}}

\newc{\sigmaplimnai}{\sigma_p^{\mathrm{lim(NaI)}}}
\newc{\sigmanlimnai}{\sigma_n^{\mathrm{lim(NaI)}}}

\newc{\ap}{a_p}			\newc{\an}{a_n}
\newc{\cp}{C_p}		\newc{\cn}{C_n}		\newc{\ca}{C_A}	
\newc{\cAp}{C_A^p}	\newc{\cAn}{C_A^n}

\newc{\rhozerothree}{\rho_{0.3}}

\newc{\sigchin}{\sigma(\chi N)}
\newc{\etazero}{\eta_0}		\newc{\deltaeta}{\Delta\eta}
\newc{\erecoil}{E_R}

\begin{document}
\topmargin-1.cm
%
\begin{titlepage}
\vspace*{-64pt}

\begin{flushright}
{SHEF-HEP/00-2,\\
MPI-PhT/2000-16,\\
May 2000}\\
\end{flushright}

\vspace{1.8cm}

\begin{center}
\large {\bf A New Model-Independent Method for Extracting Spin-Dependent
Cross Section Limits from Dark Matter Searches}\\
\vspace*{1.3cm}
\large{D.R.~Tovey$^a$ \footnote{e-mail: d.r.tovey@sheffield.ac.uk},
R.J.~Gaitskell$^b$, P.~Gondolo$^c$,}\\ \large{Y.~Ramachers$^d$ and
L.~Roszkowski$^e$} \\
\vspace*{0.4cm}
{\it\small $^a$ Department of Physics and Astronomy, University of
Sheffield, Hounsfield Rd., \\ Sheffield S3 7RH, UK.}\\
\vspace*{0.4cm}
{\it\small $^b$ Center for Particle Astrophysics, University of
California, Berkeley, \\ Berkeley, CA 94720, USA.} \\
\vspace*{0.4cm}
{\it\small $^c$ Max Planck Institute for Physics, F\o hringer Ring 6,
\\ 80805 Munich, Germany.} \\
\vspace*{0.4cm}
{\it\small $^d$Oxford University, Nuclear and Astrophysics Laboratory, Keble
Road, \\ Oxford OX1 3RH, UK.} \\
\vspace*{0.4cm}
{\it\small $^e$Department of Physics, Lancaster University, Lancaster
LA1 4YB, UK.} \\  

\end{center}

\bigskip
\begin{abstract}

A new method is proposed for extracting limits on spin-dependent
WIMP-nucleon interaction cross sections from direct detection dark
matter experiments. The new method has the advantage that the limits
on individual WIMP-proton and WIMP-neutron cross sections for a given
WIMP mass can be combined in a simple way to give a model-independent
limit on the properties of WIMPs scattering from both protons {\em
and} neutrons in the target nucleus. Extension of the technique to the
case of a target material consisting of several different species of
nuclei is discussed.
\end{abstract}

\vspace{1cm}
{\em PACS}: 95.35.+d; 12.60.Jv; 14.80.Ly

{\em Keywords}: dark matter; WIMP; neutralino; supersymmetry

\end{titlepage}

\setcounter{footnote}{0}
\setcounter{page}{0}
\newpage


\section{Introduction}

Weakly Interacting Massive Particles (WIMPs) are believed to be the
most plausible candidate for dark matter (DM) in the Universe. WIMPs
are predicted to exist in many extensions of the Standard Model of
particle physics. Most of the well-motivated WIMP candidates are
Majorana (\ie, $\bar\chi=\chi$) fermions. This in particular is often
the case in models based on supersymmetry (SUSY). Perhaps the most
popular WIMP candidate is the lightest neutralino, a superposition of
the SUSY partners of the electroweak gauge bosons (gauginos) and Higgs
particles (higgsinos). Other plausible WIMP candidates for cold DM
(CDM), the axino and gravitino (fermionic SUSY partners of the axion
and graviton, respectively) are also Majorana particles.

In the case of non-relativistic Majorana WIMPs, their predicted
elastic scattering couplings to atomic nuclei are effectively of two
types~\cite{wg}. In scalar, or spin-independent (SI) interactions the
WIMP coupling is proportional to the mass of the  nucleus. In the axial,
or spin-dependent (SD) case, the coupling is proportional to the spin
of the nucleus.

Direct searches for galactic halo WIMPs through their elastic
scattering off target nuclei are currently being carried out by
several groups. In the absence of a positive signal it is the aim of
these experiments to set limits on the properties of WIMP dark matter
independently of its precise composition. This is accomplished by
setting limits on the cross sections for SI and SD interactions
between WIMPs and target nuclei as functions of WIMP mass. To enable
comparison with results from other experiments (which may use
different target nuclei) one often translates these limits to bounds
on the WIMP-proton cross section. The current procedure used for the
conversion is relatively straightforward in the case of SI
interactions but becomes problematic in the case of SD interactions.

The reason for this is that limits on spin-dependent WIMP-proton
scattering cross sections contain considerable dependence on a
particular WIMP composition (\eg, gaugino-like versus higgsino-like
neutralino WIMP). Since the spin of target nuclei is
carried both by constituent protons and neutrons, when converting to a
WIMP-proton cross section a value for the ratio of the WIMP-proton and
WIMP-neutron cross sections must be assumed. But in many cases this
ratio can vary significantly depending on the assumed type of WIMP.
In the particular case of a predominantly gaugino neutralino WIMP this
ratio varies by several orders of magnitude~\cite{ls}. As a result,
current experimental limits on the WIMP-proton cross section for SD
interactions are fraught with potentially significant WIMP-type
dependence.

In this Letter, we introduce an alternative method for deriving limits
on spin-dependent WIMP-nucleon interactions. This new procedure allows
one to extract experimental limits in a (Majorana) WIMP-model
independent way. First, in Sec.~2, we briefly summarise the current
practice. The new method is presented in Sec.~3 and its features are
discussed in Sec.~4.

\section{The Current Procedure}

The current procedure for calculating spin-dependent WIMP-nucleon
cross section limits from experimental data can be summarised as
follows (following Refs.~\cite{ls,jkg}). Assume a detector consists of
some species of nucleus, $^A_Z X=N$. Using the convention of
Ref.~\cite{jkg} (see Eqn.~(7.13)), the total WIMP-nucleus cross
section $\sigmaa$ can be written as
\begin{equation}
\label{eq00}
\sigmaa = 4\gfermi^2 \mu_A^2 \ca,
\end{equation}
where the WIMP-target reduced mass $\mu_A$ is given by
$m_{\chi}m_A/(m_{\chi}+m_A)$ for WIMP mass $m_{\chi}$ and target
nucleus mass $m_A$. The ``enhancement factor'' $\ca$ is given in
Eqns.~(7.14) and~(7.35) of Ref.~\cite{jkg} for SD and SI interactions,
respectively, and will also be given explicitly below.

As emphasized in Ref.~\cite{jkg}, $\sigmaa$ (called there
$\sigmazero$) is, strictly speaking, {\em not} the total cross
section. It is, by definition, the ``standard'' total WIMP-nucleus
cross section at zero momentum transfer. However, it is the quantity
that is conventionally used by experimental groups for setting limits
and we will continue calling it here the total WIMP-nucleus cross
section.

In the first step, the data is used to calculate limits $\sigmaalim$
on the SI and SD WIMP-nucleus cross sections. (If the target is made
of several species of nuclei (\eg, Na and I), the procedure is
performed for each species separately and next the limits are combined
together.)  In each case it is assumed that only the given type of
interaction (SD or SI) dominates the total cross section.

The WIMP-target cross section limit $\sigmaalim$ obtained for
target $A$ can be expressed in terms of limits on WIMP-nucleon
(\ie\ {\em free} proton or neutron) cross sections $\sigmaplima$ and
$\sigmanlima$
\begin{equation}
\label{eq0}
\sigmaplima = \sigmaalim \frac{\mu_p^2}{\mu_A^2}
\frac{1}{\ca/\cp}, \qquad
\sigmanlima = \sigmaalim \frac{\mu_n^2}{\mu_A^2}
\frac{1}{\ca/\cn},
\end{equation}
where $\mu_{p,n}$ is defined by setting $A=p$ ($A=n$) in the
expression for the reduced mass $\mu_A$ above, and similarly for
$C_{p,n}$ (see also below). This conversion is made conventionally to
the WIMP-proton cross section limit using the former expression.

The purpose of this conversion is twofold. It allows one to compare
limits derived by different experiments which use different target
materials. Second, theoretical calculations in specific (\eg, SUSY)
models give predictions for $\sigmap$ (and $\sigman$) which can be
next directly compared with experimental results.

In the SI case the conversion is straightforward. This is because now
the enhancement factor is proportional to the square of the atomic
number, $\ca=\frac{1}{\pi\gfermi^2}\left[Z f_p +(A-Z) f_n \right]^2$
where $f_{p}$ and $f_{n}$ are the effective WIMP couplings to protons
and neutrons, respectively. For Majorana WIMPs $f_p\simeq f_n$ and so
one typically has $\ca/\cp \simeq \ca/\cn \simeq A^2$ and the
conversion does not depend on the specific WIMP type. For massive
Dirac neutrino-like WIMPs $f_p \simeq 0$ and $\ca/\cp \simeq \ca/\cn
\simeq (A-Z)^2$ and again the conversion does not depend on the WIMP
model or its parameters.

In the SD case however the situation is more complex. Here the
enhancement factor\footnote{ Note that the enhancement factor $\ca$ as
defined here (and in Ref.~\cite{jkg}) differs slightly from a similar
quantity $I_A$ used in Refs.~\cite{epv,ls}: $\ca=\frac{8}{\pi} I_A$.}
is given by
\begin{equation}
\ca= \frac{8}{\pi} \left(\ap\langle S_p\rangle  + \an\langle S_n\rangle
    \right)^2 \frac{J+1}{J}, 
\label{eq0.1c}
\end{equation}
where $\ap$ and $\an$ are (WIMP-type dependent) effective WIMP-proton
and WIMP-neutron couplings and $\langle S_{p,n}\rangle=\langle
N|S_{p,n}| N\rangle $ are the expectation values of the proton and
neutron spins within the nucleus and $J$ is the total nuclear spin. In
the particular case of free nucleons one finds $C_{p,n} =
\frac{6}{\pi}\, a_{p,n}^2$ and $\sigma_{p,n} = \frac{24}{\pi}
\gfermi^2 \mu_{p,n}^2 a_{p,n}^2$ where $\langle S_p\rangle= 0.5 =
\langle S_n\rangle$ has been used.

This definition of $\ca$ normalises the spin-dependent nuclear
form-factor $F^2(q)$ used in calculating nuclear recoil energy spectra
to unity at $q=0$:
\begin{equation}
\label{eq0.2}
F^2(q) = \frac{S(q)}{S(0)},
\end{equation}
where 
\begin{eqnarray}
\label{eq0.1}
S(q) & = & a_0^2S_{00}(q) + a_1^2S_{11}(q) +a_0a_1S_{01}(q)  \\ 
     & = & \left(\ap+\an\right)^2 S_{00}(q) + \left(\ap-\an\right)^2 S_{11}(q)
           + \left(\ap^2-\an^2\right) S_{01}(q). \label{eq0.1b}
\end{eqnarray}
Here the $S_{ij}$ are the respective isoscalar, isovector and
interference term form-factors for nucleus $N$ (assumed to be known
from nuclear calculations) and $a_0=\ap+\an$ and $a_1=\ap-\an$ are
isoscalar and isovector coefficients. Using this definition most of
the WIMP model-dependencies in $S(q)$ are absorbed into $\ca$, leaving
$F^2(q)$ and hence the shape of recoil energy spectra relatively
model-independent. Recent calculations~\cite{rd} have shown that
$F^2(q)$ still contains some residual model-dependencies owing to
differences in the $q$-dependence of the $S_{ij}$ form-factors. These
differences are however small for most nuclei of interest, including
Na, I and F~\cite{rpriv}, and will henceforth be neglected.

It is clear from Eqn.~(\ref{eq0.1c}) that, in the SD case, converting
WIMP-target cross section limits $\sigmaalim$ to the WIMP-proton cross
section $\sigmaplima$ becomes problematic. This is because the
enhancement factor $\ca$ now receives contributions from both proton
and neutron terms. As we will argue below, due to the presence of
WIMP-dependent coefficients $a_{p,n}$, both of these contributions can
be of comparable order and even of opposite sign. As a result, the
ratio $\ca/\cp$ in Eqn.~(\ref{eq0}), and hence the derived limit
$\sigmaplima$ will depend on the assumed type of WIMP.

In the early (single-particle or odd-group model) calculations the
nuclear spin was always assumed to be dominated either by the proton
or by the neutron term in Eqn.~(\ref{eq0.1c}). Thus when dealing with
odd-proton targets, such as Na or I, the $\ap^2$ in the expression for
$\ca$ was conveniently cancelled by the $\ap^2$ in the analogous
expression for the WIMP-proton cross section enhancement factor
$\cp$. Model dependencies were thus eliminated.

\begin{figure}
\begin{center}
\epsfig{file=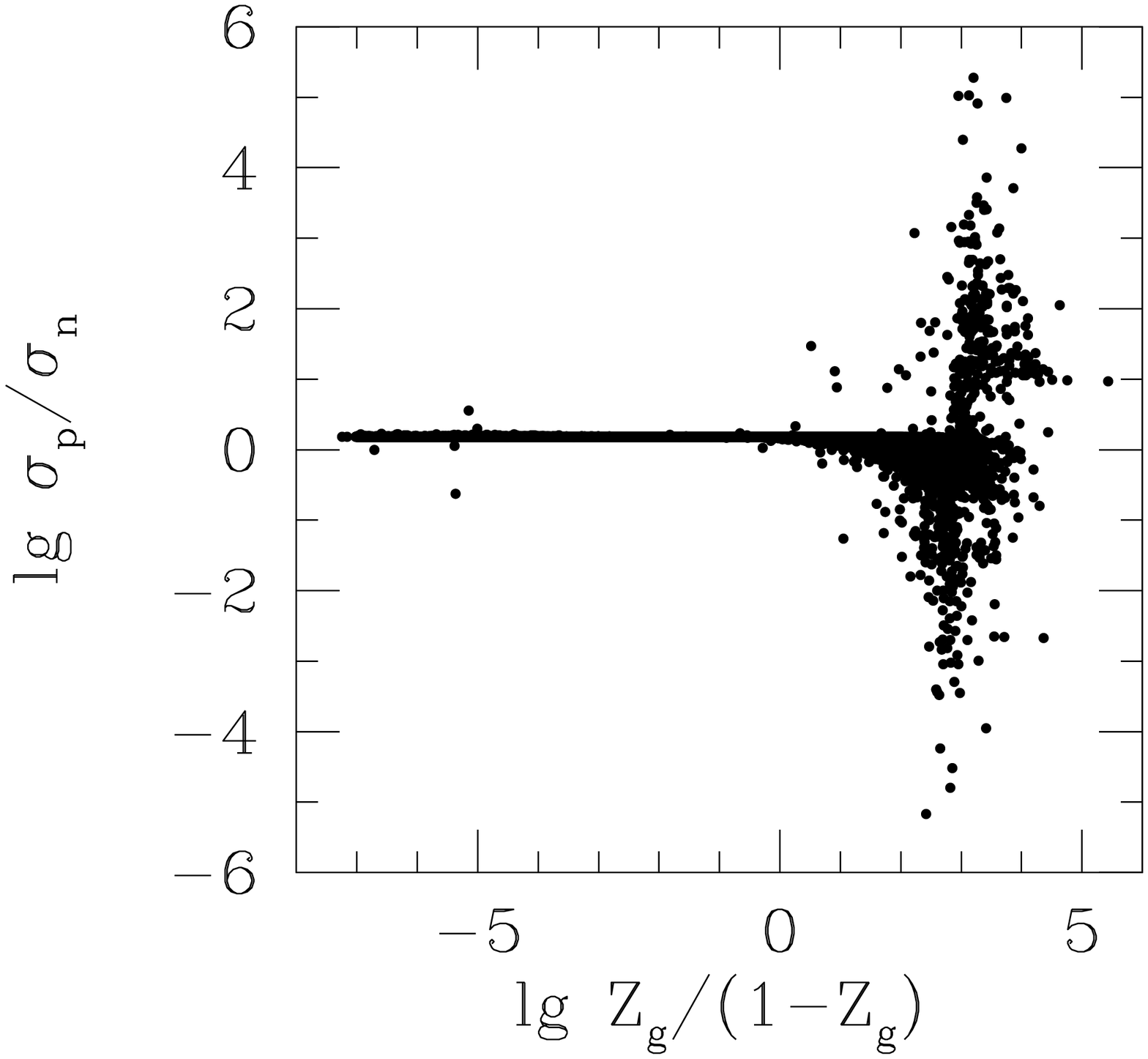,height=5.0in}
\caption{\label{fig0}}
\end{center}
\end{figure}

The conversion is however complicated when dealing with odd-neutron
targets because of the factor $(\ap/\an)^2$ remaining in the
expression for $\sigmaplima$ in Eqn.~(\ref{eq0}). This ratio is not
guaranteed to be constant and in general is WIMP model-dependent. In
the important case of SUSY neutralino WIMPs early estimates of the
$\Delta q$ values used in calculating $\ap$ and $\an$ were
nevertheless such that the ratio $\sigmap/\sigman=(\ap/\an)^2$ was
always of order unity, independent of the neutralino
composition. Later estimates of $\Delta q$~\cite{ls} have however
shown that although this is still true for predominantly higgsino
neutralinos, the ratio for gaugino neutralinos is highly SUSY
model-dependent and can vary by several orders of magnitude (as
demonstrated by Fig.~\ref{fig0} for models from the database described
in Refs.~\cite{gondx,gondy}). Gaugino-like neutralinos as DM WIMPs are
strongly favoured by a combination of naturalness and cosmological
arguments~\cite{ros} in which case this problem becomes particularly
acute.

A still further problem arises when using more recent shell-model
calculations for $\langle S_p\rangle $ and $\langle S_n\rangle
$~\cite{jkg,rd}. These indicate non-zero contributions to the nuclear
spin from {\em both} protons {\em and} neutrons, and in this case the
$\ap^2$ factor cannot be cancelled from even odd-proton nuclei. Hence
even if one of these contributions is larger than the other, as is
often the case, then the WIMP-dependent ratio of $\ap$ and $\an$ can
be such that both contributions to the cross section are
substantial. Furthermore, $\ap\langle S_p\rangle $ and $\an\langle
S_n\rangle $ can in general be of opposite sign and similar
magnitude. Hence, a more proper way of writing the enhancement factor
would be
\begin{equation}
\ca= \frac{8}{\pi} \left( |\ap\langle S_p\rangle| \pm |\an\langle
    S_n\rangle| \right)^2 \frac{J+1}{J}.
\label{eq:iapm}
\end{equation}
In general therefore, depending on the particular WIMP type, the
WIMP-target cross section can be considerably reduced relative to that
of its constituent nucleons. In these circumstances limits set by
assuming the simple case of higgsino neutralino or heavy neutrino
WIMPs (constant $(\ap/\an)$) would prove to be unduly optimistic.

\section{An Alternative Procedure}

We will now present a new method for deriving limits on spin-dependent
WIMP-nucleon interactions. The method will be free from the problems
described above and will allow one to derive experimental limits in a
way which is independent of the type of the assumed (Majorana) WIMP.

We will start by identifying the separate proton and neutron
contributions to the total enhancement factor $\ca$
\begin{equation}
\label{eq2}
\cAp = \frac{8}{\pi} \left(\ap\langle S_p\rangle\right)^2\frac{J+1}{J},
\qquad
\cAn = \frac{8}{\pi} \left(\an\langle S_n\rangle\right)^2\frac{J+1}{J}
\end{equation}
In light of Eqn.~(\ref{eq:iapm}), this gives $\ca = \left( \sqrt{\cAp}
\pm \sqrt{\cAn} \right)^2$. Following Eqns.~(\ref{eq00})
and~(\ref{eq0.1c}) we thus define the proton and neutron contributions
$\sigmaap$ and $\sigmaan$ to the total cross section $\sigmaa$ as:
\begin{equation}
\label{eq9-1}
\sigmaap = 4\gfermi^2 \mu_A^2 \cAp
\qquad
\sigmaan = 4\gfermi^2 \mu_A^2 \cAn.
\end{equation}
Using Eqns.~(\ref{eq00}), (\ref{eq:iapm}) and (\ref{eq9-1}), one can
express $\sigmaa$ as
\begin{equation}
\label{eq9-2}
\sigmaa = \left(\sqrt{\sigmaap} \pm
\sqrt{\sigmaan}\right)^2.
\end{equation} 

We note that $\sigmaap$ and $\sigmaan$ are {\em not} measured cross
sections. They are nevertheless convenient auxiliary quantities which
identify separate proton and neutron contributions to the total cross
section $\sigmaa$.

We will now proceed as follows. (See the Appendix for a more rigorous
treatment.) We will first make an auxiliary assumption that $\sigmaa
\simeq \sigmaap$. In other words, we assume that the total
WIMP-nucleus cross section is dominated by the proton contribution
only. We then {\em define} the WIMP-proton cross section limit
$\sigmaplima$ corresponding to the WIMP-target $A$ cross section limit
$\sigmaalim$ as
\begin{equation}
\label{eq4.-1p}
\sigmaplima = \sigmaalim \frac{\mu_p^2}{\mu_A^2} \frac{1}{\cAp/\cp}.
\end{equation}
Analogously, the WIMP-neutron cross section limit $\sigmanlima$ is
defined by assuming that $\sigmaa \simeq \sigmaan$ and writing
\begin{equation}
\label{eq4.-1n}
\sigmanlima = \sigmaalim \frac{\mu_n^2}{\mu_A^2} \frac{1}{\cAn/\cn}.
\end{equation}

It is clear that the use of the ratios $\cAp/\cp=4/3 \langle
S_p\rangle^2 (J+1)/J$ and $\cAn/\cn=4/3 \langle S_n\rangle^2 (J+1)/J$
ensures the cancellation of the WIMP-dependent $\ap^2$ and $\an^2$
terms contained within the WIMP-target cross section $\sigmaa$ and
hence ensures WIMP model-independence. Values of $\cAp/\cp$ and
$\cAn/\cn$ for typical nuclei of interest obtained using data from
Refs.~\cite{jkg,rd,rpriv} are listed in Table~\ref{t1}.

These properties can now be used in expressing WIMP-independent
experimental limits on the WIMP-nucleus SD interaction cross section
$\sigmaalim$ in terms of $\sigmap$ and $\sigman$ which are the
quantities whose values are predicted by specific theoretical models.
If an experiment publishes the limits $\sigmaplima$ and $\sigmanlima$
then Eqn.~(\ref{eq9-2}) can be used to define a WIMP-independent
excluded region in the $\sigmap$ - $\sigman$ plane as
\begin{equation}
\label{eq5}
\left(\sqrt{\frac{\sigmap}{\sigmaplima}} \pm
\sqrt{\frac{\sigman}{\sigmanlima}}\right)^2 > 1.
\end{equation}

Because of relative sign ambiguity, the condition~(\ref{eq5}) implies
two bounds corresponding to constructive and destructive interference.
A conservative limit corresponds to the relative minus sign which
reduces the overall WIMP-target cross section. We note, however, that
in comparing with a specific theoretical (e.g. SUSY) model there will
be no sign ambiguity: the theoretical model predicts not only
$\sigmap$ and $\sigman$ but also the signs of $\ap$ and $\an$. In this
case the sign in Eqn.~(\ref{eq5}) is known and is given by the sign of
$(\ap\langle S_p\rangle) /(\an\langle S_n\rangle)$.

An alternative way of expressing the limits in this procedure is to
consider exclusion regions directly in terms of the fundamental
WIMP-nucleon coupling coefficients $\ap$ and $\an$. In this case
Eqn.~(\ref{eq5}) is replaced by
\begin{equation}
\label{eq5a}
\left(\frac{\ap}{\sqrt{\sigmaplima}} \pm
\frac{\an}{\sqrt{\sigmanlima}}\right)^2 >
\frac{\pi}{24\gfermi^2\mu_p^2},
\end{equation}
where we applied Eqn.~(\ref{eq00}) to the case of the nucleons to
obtain $\sigmap = 24\gfermi^2 \mu_p^2 \ap^2/\pi$ and $\sigman =
24\gfermi^2\mu_n^2 \an^2/\pi$. The small proton-neutron mass
difference has been ignored.

The variables $\ap$ and $\an$ can have either sign, and the relative
sign inside the square is now determined by the sign of $\langle
S_p\rangle /\langle S_n\rangle $ only. Eqn.~(\ref{eq5a}) corresponds
geometrically in the $\ap$ - $\an$ plane to excluding a region
exterior to two parallel lines whose slope has opposite sign to
$\langle S_p \rangle / \langle S_n \rangle$. There is no limit on
$\ap$ or $\an$ between these lines. This region extends to infinity in
both directions.

So far, we have presented the method for one species of target nucleus
only. A generalisation to two or more nuclei in the same target is
straightforward. Analogously to the WIMP-proton cross section limit in
the current method~\cite{ls}, the limits $\sigmaplimai$ and
$\sigmanlimai$ from different nuclei $A_i$ in the target $A$ can be
combined by using
\begin{equation}
\label{eq4}
\frac{1}{\sigmaplima} = \sum_{A_i}
\frac{1}{\sigmap^{\mathrm{lim}(A_i)}}, \qquad
\frac{1}{\sigmanlima} = \sum_{A_i}
\frac{1}{\sigman^{\mathrm{lim}(A_i)}}.
\end{equation}

It should be noted that when calculating the combined excluded region
in $\sigmap$ - $\sigman$ parameter space for two or more nuclei in the
same target material it would be incorrect to use Eqn.~(\ref{eq5})
with the combined limits $\sigmaplima$ and $\sigmanlima$ calculated
using Eqn.~(\ref{eq4}). The correct approach is to use instead the
generalisation of Eqn.~(\ref{eq5}) given by
\begin{equation}
\label{eq6}
\sum_{A_i} \left(\sqrt{\frac{\sigmap}{\sigmap^{\mathrm{lim}(A_i)}}} \pm
\sqrt{\frac{\sigman}{\sigman^{\mathrm{lim}(A_i)}}}\right)^2 > 1.
\end{equation}  

A similar generalisation can also be applied to Eqn.~(\ref{eq5a}). In
this case for given $\sigmaplima$ and $\sigmanlima$ optimum limits on
the coupling coefficients $\ap$ and $\an$ could be obtained by using
two different target nuclei with $\langle S_p\rangle /\langle
S_n\rangle $ of opposite sign, such as are found in NaCl or NaF. The
allowed region would then lie inside the intersection of the two bands
of opposite slope.

As a practical illustration, we consider limits from a synthetic data
set (assumed to be from a NaI detector) using both the current and
the proposed techniques. The data set consists of recoil energy
dependent event rate limits which have been converted to WIMP mass
dependent target nucleus cross section limits using nuclear kinematics
and Eqn.~(\ref{eq0.2}) as described in Ref.~\cite{ls}.

In Fig.~\ref{fig1-1}(a) we plot the limits calculated using the
current technique (\ie\ using the expression for $\sigmaplima$ given
in Eqn.~(\ref{eq0}) with $\ca$ from
Eqn.~(\ref{eq0.1c})). We do this for three different neutralino WIMP
compositions (\ie\ different ratios $\ap/\an$ from
Fig.~\ref{fig0}). This results in overall (combined Na and I) limits
$\sigmaplimnai$ which, for a given WIMP mass but different type, can
be different by almost two orders of magnitude : a thick solid curve
corresponds to the limit for the higgsino WIMP ($\ap/\an\sim1.5$),
which is currently commonly assumed, while the two thick dash-dotted
lines correspond to the two gaugino WIMP cases giving the extremal
limits for WIMPs of mass 100 GeV/c$^2$. These two curves illustrate the
considerable effects of destructive (upper curve) and constructive
(lower curve) interference between proton and neutron contributions to
the nuclear spin.

Results obtained using the proposed technique are presented in
Figs.~\ref{fig1-1}(b) and (c). We plot the limits $\sigmaplima$ and
$\sigmanlima$ calculated using Eqns.~(\ref{eq4.-1p}) - (\ref{eq4.-1n})
and hence avoiding the assumption of a specific WIMP composition. We
can clearly see that the limits in windows (a), (b) and (c) differ
considerably. In all the cases at small WIMP masses the best limits
are provided by Na due to its slowly-varying form-factor
$F^2(q)$. However, at larger masses iodine provides a better limit due
to its larger value for $\cAp/\cp$ (Table~\ref{t1}). The spin of both
nuclei is carried predominantly by protons and so for this target the
limits $\sigmaplima$ are in general superior to $\sigmanlima$.

The individual limits $\sigmaplima$ and $\sigmanlima$ presented in
Figs.~\ref{fig1-1} (b) and (c) contain all the information which
allows one to draw, for a given WIMP mass, exclusion regions in the
$\sigmap$ - $\sigman$ plane by using Eqn.~(\ref{eq6}). This is shown
in Fig.~\ref{fig1}.  We use the same data set as before and set a WIMP
mass of 100 GeV/c$^2$. Figs.~\ref{fig1}(a) and (b) correspond to minus
and plus signs respectively in Eqn.~(\ref{eq6}), although in the
absence of knowledge of this sign the former plot gives the more
conservative limits.

Finally we note that, in the presented technique, one can incorporate
limits $\sigma_{SI}^{\mathrm{lim}(A_i)}$ set on the spin-independent
WIMP-nucleon cross section $\sigma_{SI}$ by constituent nuclei $A_i$
using the technique of Sec.~1:
\begin{equation}
\label{eq7}
\sum_{A_i} \left\lgroup
\left(\sqrt{\frac{\sigmap}{\sigmap^{\mathrm{lim}(A_i)}}} \pm
\sqrt{\frac{\sigman}{\sigman^{\mathrm{lim}(A_i)}}}\right)^2 +
\left(\frac{\sigma_{SI}}{\sigma_{SI}^{\mathrm{lim}(A_i)}}\right)\right\rgroup
> 1.
\end{equation}

In summary, in contrast to the currently used practice of using only
the WIMP-proton cross section $\sigmap$ for presenting the SD
WIMP-nucleus cross section limits, this alternative method effectively
uses both the WIMP-proton and WIMP-neutron cross sections $\sigmap$
{\em and} $\sigman$. This ensures that the WIMP-model dependence of
the experimental SD cross section limits is removed. The method
applies to targets with one or more species of nuclei in the target
and allows unambiguous comparisons with theoretical predictions.

\section{Discussion and Conclusions}

In contrast to the currently used procedure, the new technique makes
it possible for a given experiment to set limits on the spin-dependent
WIMP-nucleon cross section in a WIMP-independent way. Another
significant advantage of the new technique is that it makes the direct
comparison of experimental results with theoretical predictions
possible.  Given the WIMP-proton and WIMP-neutron exclusion plots from
a particular experiment, one can determine whether a specific choice
of parameters is allowed or excluded for a given WIMP mass simply by
using Eqn.~(\ref{eq7}). In fact, all the information needed to allow
or exclude any WIMP candidate (like SUSY neutralino WIMPs with
arbitrary composition, non-SUSY heavy neutrino, etc.) can be presented
in only three figures (e.g. Fig.~\ref{fig1-1}(b) and
Fig.~\ref{fig1-1}(c) plus one SI cross section limit plot). No
additional plots similar to Fig.~\ref{fig1} need be drawn to determine
the exclusion region for a particular WIMP mass.

It is not surprising that dark matter experiments using odd-proton
targets will generally give very good WIMP-proton cross section limits
and rather weak WIMP-neutron cross section limits for the same
WIMP-target cross section limits. Just the opposite will be true with
experiments using odd-neutron targets. This is what one should in fact
expect: odd-proton and odd-neutron targets are effectively measuring
two quite different and unrelated quantities ($\sigmap$ and
$\sigman$). It is only in the particular case of higgsino neutralino
and heavy neutrino WIMPs~\cite{ls} that the limits from such targets
calculated using the new technique are still comparable, by using a
constant value for $\ap/\an\sim1.5$ to combine $\sigmaplima$ and
$\sigmanlima$ into a single limit on the WIMP-proton cross
section. These limits are equivalent to those obtained using the
current technique (\ie\ Fig.~\ref{fig1-1}(a)).  In any case,
experiments may find it useful to publish such limits in addition to
the $\sigmap$ and $\sigman$ limit curves calculated using the new
technique (\ie\ Fig.~\ref{fig1-1}(b) and Fig.~\ref{fig1-1}(c)).

In the event of a discovery of spin-dependent WIMP-nucleon
interactions it is likely that a different procedure from that
described above would be required to analyse the data. In this case a
fit would likely be performed to the observed nuclear recoil energy
spectrum, with parameters such as the WIMP mass and cross sections on
each target nucleus being considered to be free. This would in turn
define an allowed region in $\sigmap$ - $\sigman$ parameter space
for each allowed WIMP mass and it would be this region which could be
compared with theoretical predictions.

In conclusion, a new technique has been presented for deriving
spin-dependent WIMP-nucleon cross section limits from direct detection
dark matter experiments. This technique retains the attractive
features of the current procedure, including the ability to combine
limits from individual target nuclei. The new technique, however, has
the additional advantage of allowing the calculation of WIMP-nucleon
spin dependent cross section limits in a WIMP-independent manner and
of making it possible to compare experimental results with theoretical
predictions.
\newpage

\section*{Appendix}

Here we present the basic formalism behind the alternative method
presented in Sec.~3. For the sake of completeness, we will include
here some expressions given already there.

The separate proton and neutron contributions to the total
enhancement factor $\ca$ given in Eqn.~(\ref{eq0.1c}) are given in
Eqns.~(\ref{eq2})
\[
\cAp = \frac{8}{\pi} \left(\ap\langle S_p\rangle\right)^2\frac{J+1}{J},
\qquad
\cAn = \frac{8}{\pi} \left(\an\langle S_n\rangle\right)^2\frac{J+1}{J}
\]
so that, from Eqn.~(\ref{eq:iapm}), $\ca = \left( \sqrt{\cAp} \pm
\sqrt{\cAn} \right)^2$. Following Eqns.~(\ref{eq00})
and~(\ref{eq0.1c}) the proton and neutron contributions $\sigmaap$ and
$\sigmaan$ to the total cross section $\sigmaa$ are defined as
(Eqns.~(\ref{eq9-1})):
\[
\sigmaap = 4\gfermi^2 \mu_A^2 \cAp
\qquad
\sigmaan = 4\gfermi^2 \mu_A^2 \cAn.
\]
Using Eqns.~(\ref{eq00}), (\ref{eq:iapm}) and (\ref{eq9-1}), one can
then write $\sigmaa = \left(\sqrt{\sigmaap} \pm
\sqrt{\sigmaan}\right)^2$ (Eqn.~(\ref{eq9-2})).

Applying Eqn.~(\ref{eq00}) to protons and neutrons respectively and
comparing with Eqns.~(\ref{eq9-1}) we find the following relations
\begin{equation}
\label{eq:3ap}
\sigmap = \sigmaap \frac{\mu_p^2}{\mu_A^2} \frac{\cp}{\cAp}, \qquad
\sigman = \sigmaan \frac{\mu_n^2}{\mu_A^2} \frac{\cn}{\cAn}.
\end{equation}
It is clear that the use of the ratios $\cp/\cAp$ and $\cn/\cAn$
ensure the cancellation of the $\ap^2$ and $\an^2$ terms contained
within WIMP-target cross section $\sigmaa$ and hence ensure WIMP
model-independence. Unfortunately, the quantities $\sigmaap$ and
$\sigmaan$ cannot be measured directly. We will therefore make
independently the assumptions that
\begin{equation}
\label{eq9:4ap}
\sigmaa \simeq \sigmaap, \qquad
\sigmaa \simeq \sigmaan,
\end{equation}
in which case Eqns.~(\ref{eq:3ap}) become
\begin{equation}
\label{eq9:5ap}
\sigmap \rightarrow \sigmapa = \sigmaa \frac{\mu_p^2}{\mu_A^2}
\frac{\cp}{\cAp}, \qquad
\sigman \rightarrow \sigmana = \sigmaa \frac{\mu_n^2}{\mu_A^2}
\frac{\cn}{\cAn}.
\end{equation}
In other words $\sigmapa$ and $\sigmana$ are the WIMP-proton and
WIMP-neutron cross sections derived from the WIMP-target cross section
$\sigmaa$ by assuming that it is dominated by its proton and neutron
contributions respectively. These new quantities are related to
$\sigmap$ ($\sigman$) and $\sigmaap$ ($\sigmaan$) by
\begin{equation}
\label{eq9:6ap}
\frac{\sigmap}{\sigmapa} = \frac{\sigmaap}{\sigmaa}, \qquad
\frac{\sigman}{\sigmana} = \frac{\sigmaan}{\sigmaa}.
\end{equation}
Hence from Eqn.~(\ref{eq9-2}),
\begin{equation}
\left(\sqrt{\frac{\sigmap}{\sigmapa}} \pm
\sqrt{\frac{\sigman}{\sigmana}}\right)^2 = 1.
\end{equation}

Experiments set limits $\sigmaalim$ on $\sigmaa$, while $\sigmap$ and
$\sigman$ are quantities whose values are predicted by theoretical
models. It is our goal to set limits on $\sigmap$ and $\sigman$ by
using the measured values of $\sigmaalim$. In direct analogy with
Eqn.~(\ref{eq0}) we then define the WIMP-proton cross section limit
$\sigmaplima$ corresponding to the WIMP-target $A$ cross section limit
$\sigmaalim$ as
\[
\sigmaplima = \sigmaalim \frac{\mu_p^2}{\mu_A^2} \frac{\cp}{\cAp},
\qquad
\sigmanlima = \sigmaalim \frac{\mu_n^2}{\mu_A^2} \frac{\cn}{\cAn},
\]
which are Eqns.~(\ref{eq4.-1p}) - (\ref{eq4.-1n}) in Sec.~3.  If an
experiment publishes the total SD WIMP-target cross section limit
$\sigmaalim$ in terms of the corresponding limits $\sigmaplima$ and
$\sigmanlima$ then a WIMP-independent excluded region in the $\sigmap$
- $\sigman$ plane will be given by the condition~(\ref{eq5})
\[
\left(\sqrt{\frac{\sigmap}{\sigmaplima}} \pm
\sqrt{\frac{\sigman}{\sigmanlima}}\right)^2 > 1.
\]

\section*{Acknowledgments}
DRT wishes to thank J.D. Lewin for extensive discussions and PPARC for
support. We would like to thank CERN and everyone involved with the
organisation of the Dark Matter Tools '99 Workshop where the bulk of
this work was carried out.

\newpage

\section*{Tables}

Table 1: Values of $\langle S_p\rangle $, $\langle S_n\rangle $,
$\cAp/\cp$ and $\cAn/\cn$ for various nuclei. Values of $\langle
S_p\rangle $ and $\langle S_n\rangle $ for Na, Te, I and Xe are taken
from Ref.~\cite{rd}. Values for F are taken from
Ref.~\cite{rpriv}. All others are from the review of Ref.~\cite{jkg}
and the references contained therein.


\section*{Figures}
Figure 1: The ratio $\sigmap/\sigman$ plotted against neutralino
composition $Z_g/(1-Z_g)$ for models from the database built up in
Refs.~\cite{gondx,gondy}. Here $Z_g$ is the gaugino fraction and
$(1-Z_g)$ is the higgsino fraction.  Small values of $Z_g/(1-Z_g)$
correspond to predominantly higgsino neutralinos while large values
correspond to predominantly gaugino neutralino. In all models plotted,
the neutralino is a good dark matter candidate (\ie ~its relic density
is in the range $0.025 < \Omega_\chi h^2 < 1$). The neutralino-proton
and neutralino-neutron cross sections $\sigma_p$ and $\sigma_n$ are
calculated as in Ref.~\cite{gondx}. \\
\newline
Figure 2: Exclusion regions for simulated data from a NaI
detector. Figure (a) shows in the current technique the limits
$\sigmaplima$ calculated using Eqn.~(\ref{eq4}) for the case of three
different neutralino WIMPs.  The thick solid curve is the combined
limit for a higgsino-like WIMP (corresponding to $\ap/\an\sim1.5$),
which is currently commonly assumed. The two thick dash-dotted curves
are the combined limits calculated for two different gaugino
neutralino cases assuming either destructive (upper curve) or
constructive (lower curve) interference respectively between proton
and neutron contributions to the nuclear spin.  Also shown are the
individual limits from Na (dashed) and I (dotted) nuclei contributing
to the combined limit on the higgsino-like WIMP.\\ Figures (b) and (c)
show the limits $\sigmaplima$ and $\sigmanlima$ respectively
calculated from the same data using Eqns.~(\ref{eq4.-1p}) -
(\ref{eq4.-1n}) in the framework of the new technique. Dashed (dotted)
curves again correspond to individual limits from Na (I) nuclei and
thick solid curves to combined limits obtained using
Eqn.~\ref{eq4}. The limits are neutralino WIMP type independent.  \\
\newline

\begin{figure}
\begin{center}
\epsfig{file=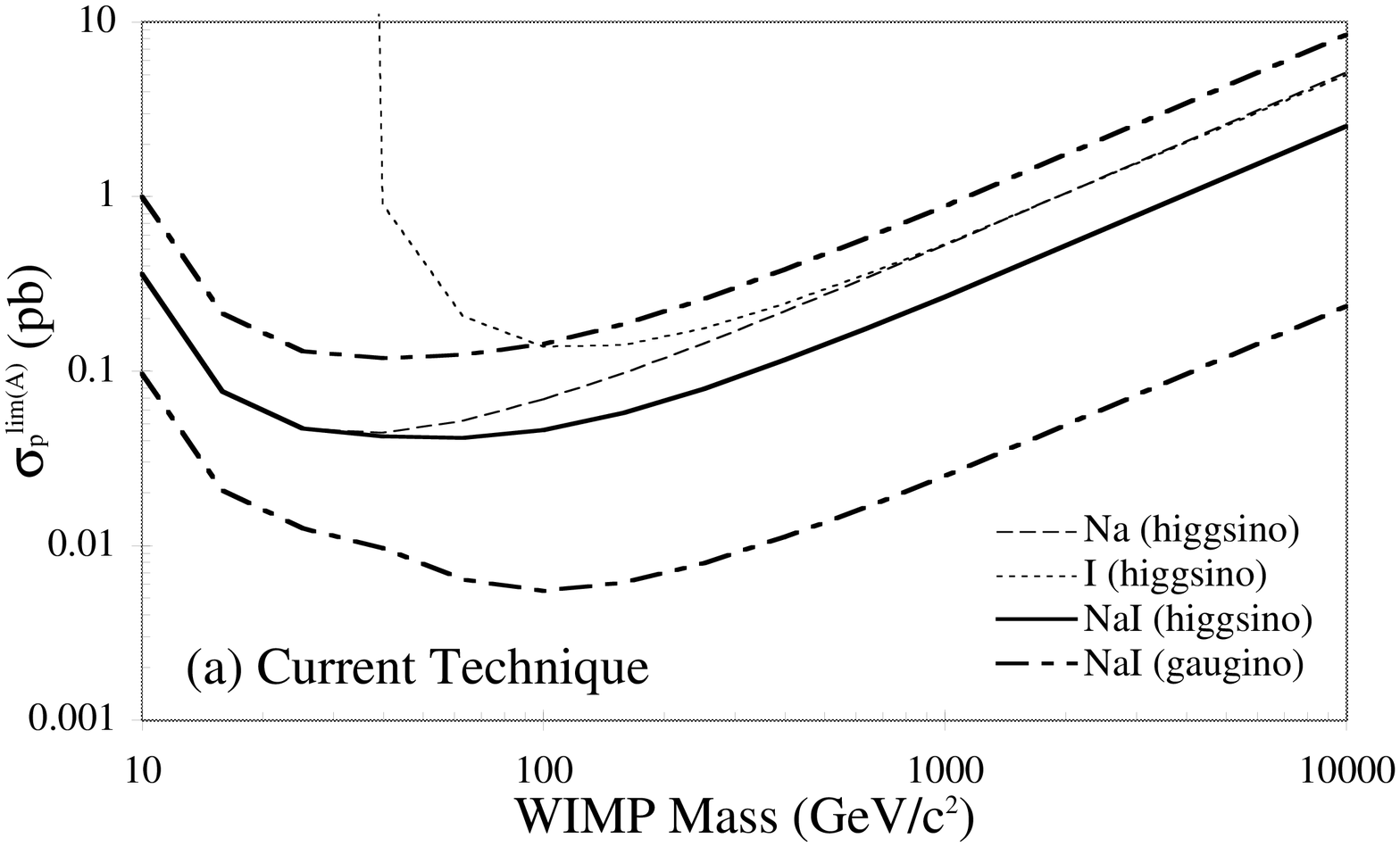,height=2.5in}
\epsfig{file=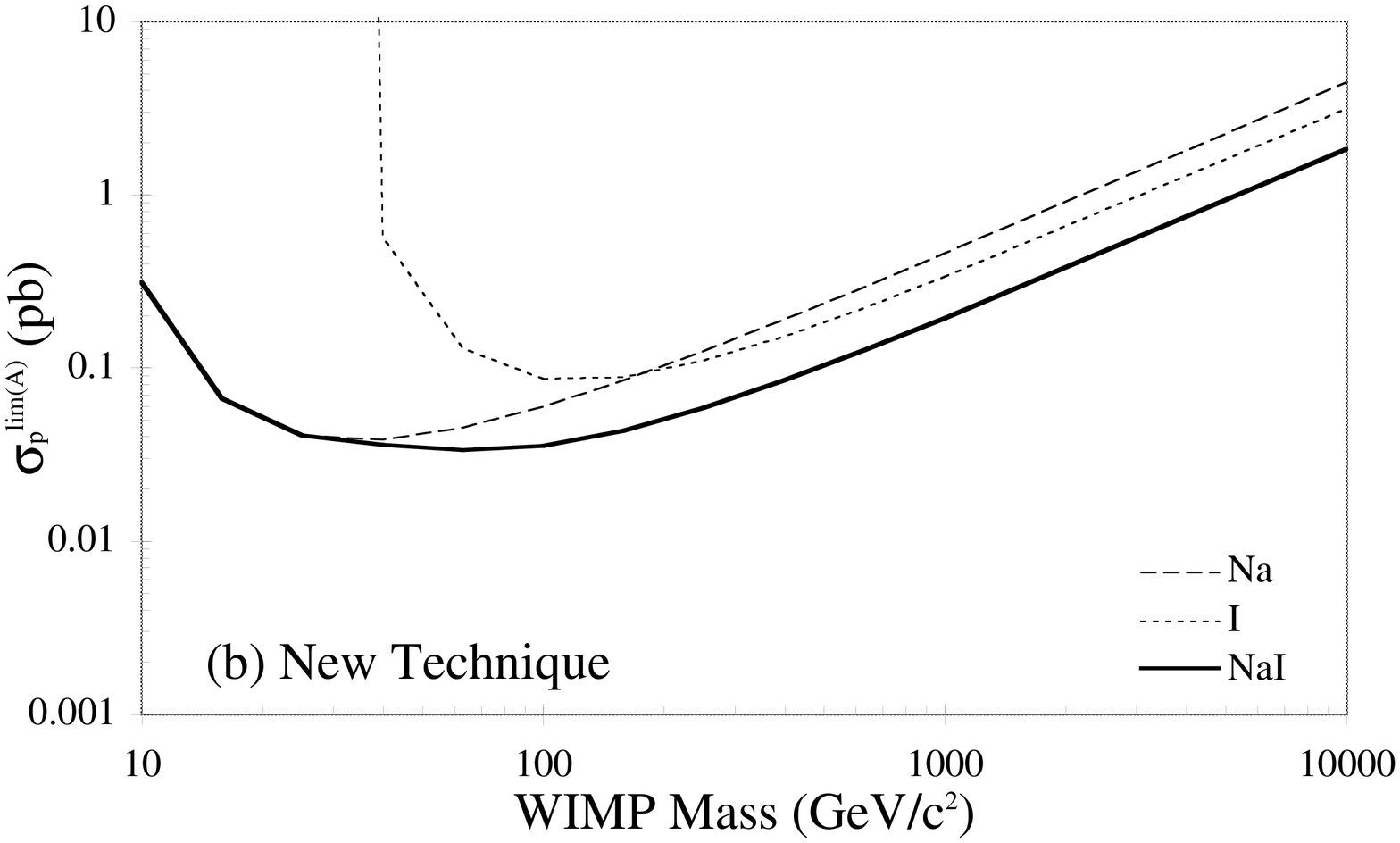,height=2.5in}
\epsfig{file=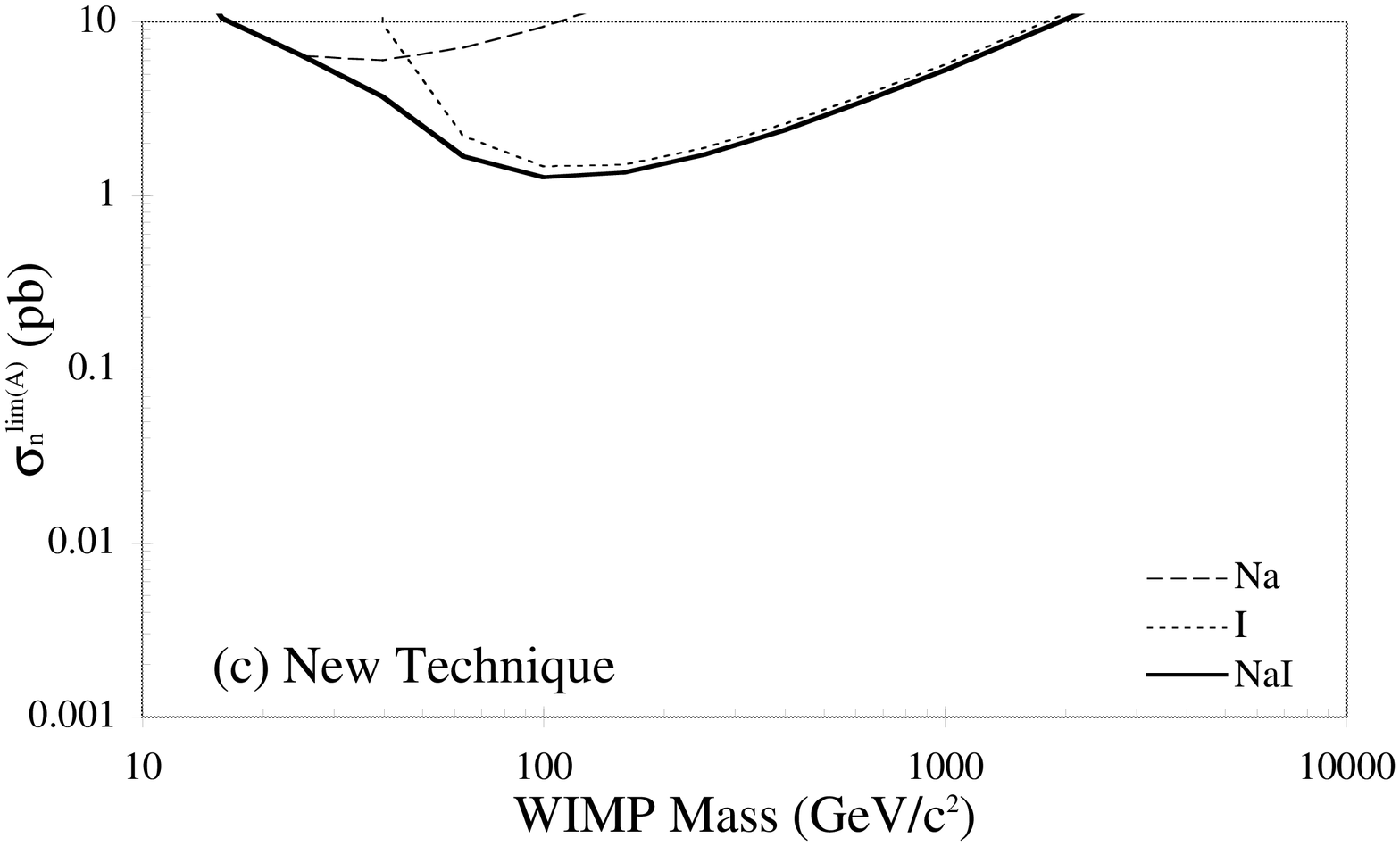,height=2.5in}
\caption{\label{fig1-1}}
\end{center}
\end{figure}

\begin{figure}
\begin{center}
\epsfig{file=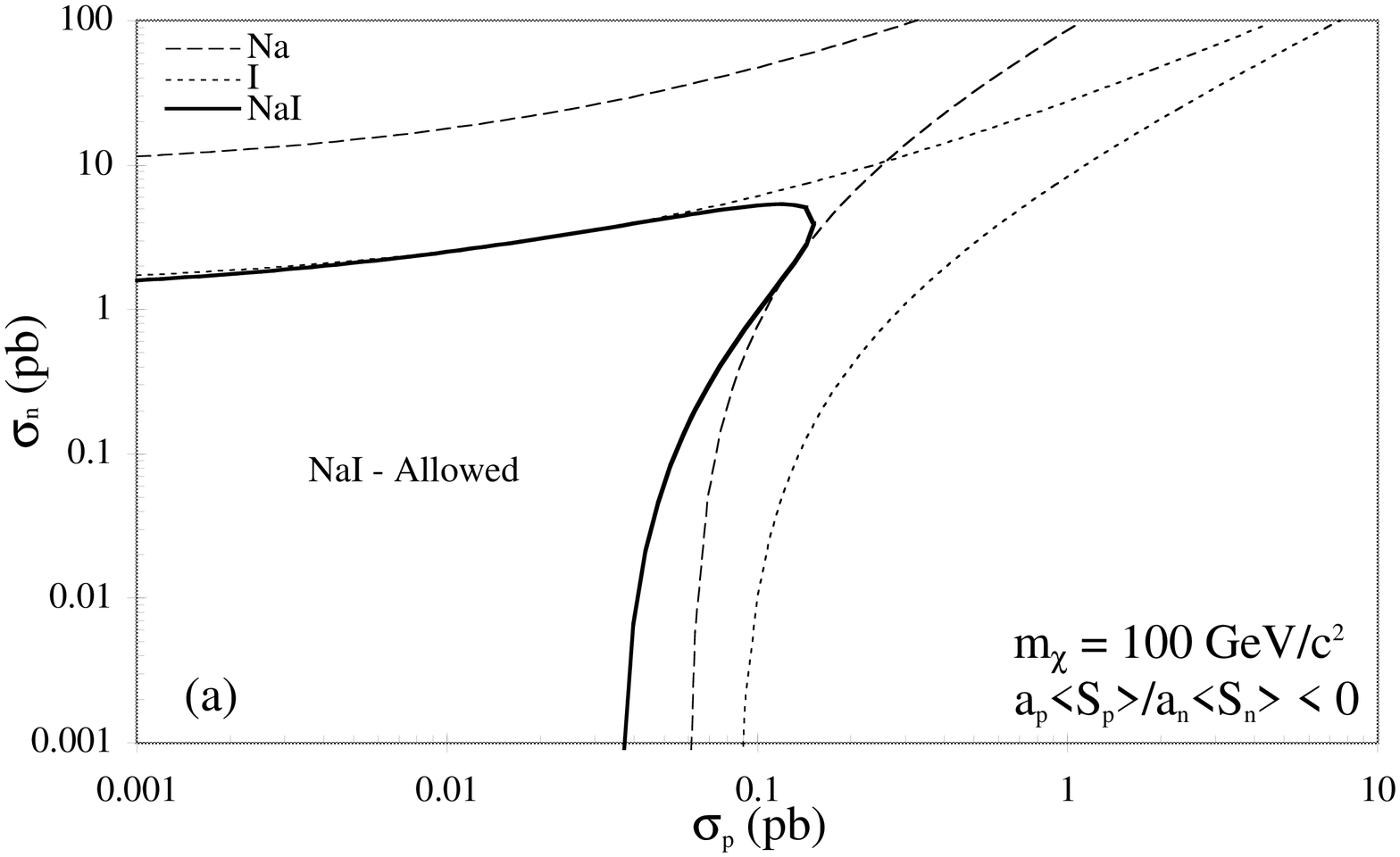,height=3.0in}
\epsfig{file=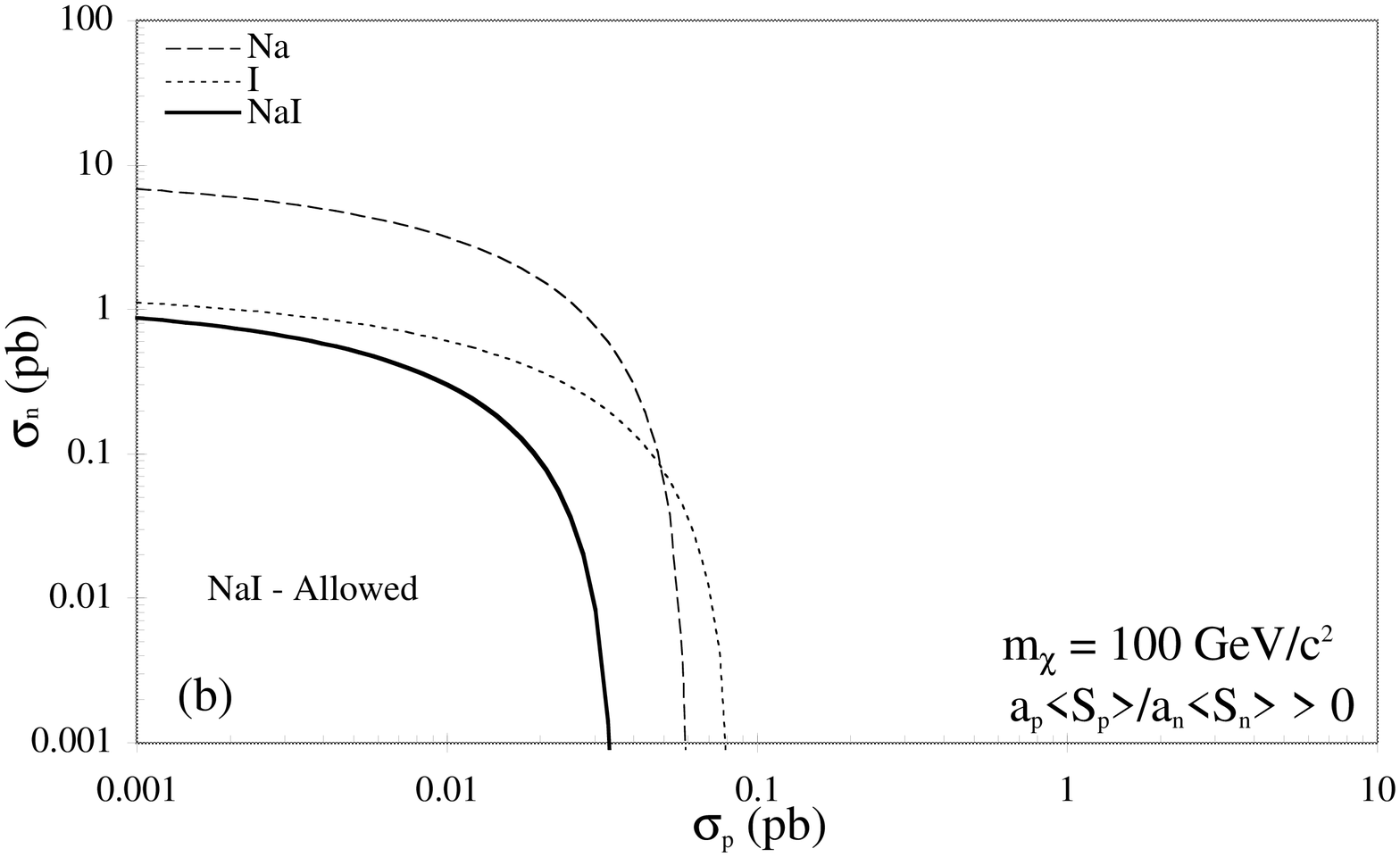,height=3.0in}
\caption{\label{fig1}}
\end{center}
\end{figure}

Figure 3: Exclusion regions in $\sigmap$ - $\sigman$ plane are plotted
in the case of 100 GeV/c$^2$ WIMPs calculated from the data of
Fig.~\ref{fig1-1}(b) and (c) using Eqn.~\ref{eq6}. Figure (a) is for
the conservative case of destructive interference ($(\ap\langle
S_p\rangle)/(\an\langle S_n\rangle)< 0$) while Figure (b) is for the
case of constructive interference ($(\ap\langle
S_p\rangle)/(\an\langle S_n\rangle)> 0$). Dashed (dotted) curves
correspond to limits from Na (I) alone while the full thick curves
show the combined limits. Note that in Figure (a) no limit can be set
by any one nucleus when $\sigmap/\sigmaplima = \sigman/\sigmanlima$
due to destructive interference between the proton and neutron
contributions. Combination of limits from two nuclei with different
$\sigmaplima/\sigmanlima$ allows such a limit to be set however.
\newpage 

\begin{table}[thb]
\begin{center}
\begin{tabular}{||c|l|c|c|r|r|l|l||}				\hline\hline
Nucleus &$Z$ &Odd Nucleon &$J$ &$\langle S_p\rangle $ &$\langle S_n\rangle $
&$\cAp/\cp$ &$\cAn/\cn$ \\ \hline\hline
$^{19}$F	&9	&p	&1/2	&0.477	&-0.004	&9.10$\times 10^{-1}$	&6.40$\times 10^{-5}$	\\ \hline
$^{23}$Na	&11	&p	&3/2	&0.248	&0.020	&1.37$\times 10^{-1}$	&8.89$\times 10^{-4}$	\\ \hline
$^{27}$Al	&13	&p	&5/2	&-0.343	&0.030	&2.20$\times 10^{-1}$	&1.68$\times 10^{-3}$	\\ \hline
$^{29}$Si	&14	&n	&1/2	&-0.002	&0.130	&1.60$\times 10^{-5}$	&6.76$\times 10^{-2}$	\\ \hline
$^{35}$Cl	&17	&p	&3/2	&-0.083	&0.004	&1.53$\times 10^{-2}$	&3.56$\times 10^{-5}$	\\ \hline
$^{39}$K	&19	&p	&3/2	&-0.180	&0.050	&7.20$\times 10^{-2}$	&5.56$\times 10^{-3}$	\\ \hline
$^{73}$Ge	&32	&n	&9/2	&0.030	&0.378	&1.47$\times 10^{-3}$	&2.33$\times 10^{-1}$	\\ \hline
$^{93}$Nb	&41	&p	&9/2	&0.460	&0.080	&3.45$\times 10^{-1}$	&1.04$\times 10^{-2}$	\\ \hline
$^{125}$Te	&52	&n	&1/2	&0.001	&0.287	&4.00$\times 10^{-6}$	&3.29$\times 10^{-1}$	\\ \hline
$^{127}$I	&53	&p	&5/2	&0.309	&0.075	&1.78$\times 10^{-1}$	&1.05$\times 10^{-2}$	\\ \hline
$^{129}$Xe	&54	&n	&1/2	&0.028	&0.359	&3.14$\times 10^{-3}$	&5.16$\times 10^{-1}$	\\ \hline
$^{131}$Xe	&54	&n	&3/2	&-0.009	&-0.227	&1.80$\times 10^{-4}$	&1.15$\times 10^{-1}$	\\ \hline\hline
\end{tabular}
\vspace{1.0 cm}
\caption{\label{t1}{\it }}
\end{center}
\end{table}
 
\newpage

\newpage

\newpage


\begin{thebibliography}{999}
\bibitem{wg} M.W. Goodman, E. Witten, {\em Phys. Rev.} {\bf D31}
(1985) 3059-3063.
\bibitem{ls} J.D. Lewin, P.F. Smith, {\em Astropart. Phys.} {\bf 6}
(1996) 87-112.
\bibitem{jkg} G. Jungman, M. Kamionkowski, K. Griest, {\em Phys. Rep.}
{\bf 267} (1996) 195-373.
\bibitem{epv} J. Engel, S. Pittel, P.Vogel, {\em Int. J. Mod. Phys.}
{\bf E1} (1992) 1-37.
\bibitem{rd} M.T. Ressell, D.J. Dean, {\em Phys. Rev.} {\bf C56}
(1997) 535-546.
\bibitem{rpriv} M.T. Ressell, {\em Private Communication}.
\bibitem{gondx} L. Bergstr\"om, P. Gondolo, {\em Astropart. Phys.}
{\bf 5} (1996) 263-278.
\bibitem{gondy} J. Edsj\"o, P. Gondolo, {\em Phys. Rev.} {\bf D56}
(1997) 1879-1894; L. Bergstr\"om, P. Ullio, {\em Nucl. Phys.} {\bf
B504} (1997) 27-44; E.A. Baltz, J. Edsj\"o, {\em Phys. Rev.} {\bf
D5902} (1999) 3511-3524. 
\bibitem{ros} L. Roszkowski, {\em Phys. Lett.} {\bf B262} (1991) 59-67.
\end{thebibliography}
\end{document}